# Epigenetic effects of cytosine derivatives are caused by their tautomers in Hoogsteen base pairs

Denis Semyonov.
E-mail address: dasem@mail.ru

***Summary***. *Deoxycitidine in solution exists as two tautomers one of which is an "uncanonical" imino one. The latter can dominate with such derivatives as 5-methyl, 5-hydroxymethyl- and 5-formylcytosine. The imino tautomer potentially is able to form a hoosteen GC base pair. To detect such pair, it is suggested to use 1H15N NMR. Formation of GC-Hoogsteen base pair with imino tautomer of cytosine can be a reason for epigenetic effects of 5-methyl- and , 5-hydroxymethylcytosine.*

Introduction. Tautomers of heterocyclic bases of nucleic acids are formed via intramolecular transfer of a single proton, and cytosine can exist as the amino- (canonical) and imino- (rare) tautomer. NMR data on deoxycitidine solution (unpublished results) show the presence of both tautomers in comparable amounts. Derivatives of cytosine, such as 5-methylcytosine and 5-hydroxymethylcytosine, are expected to form the imino tautomer with higher probability. It is known that these derivatives actively participate in regulation of genes expression since DNA regions containing derivatized cytosines are recognized by special regulatory proteins. Participation in regulation of genes is considered as epigenetic effect of methylation and hydroxymethylation of cytosines. Tautomeric properties of cytosine derivatives can underlie their epigenetic effects.

The imino tautomer, which potentially can form a GC-Hoogsteen base pair. The ability of the imino-tautomer to form such pair has not been observed so far, and it is possible that it explains epigenetic effects of cytosine derivatives

1. It is known that deoxycitidine in solution at neutral pH exists in two tautomeric forms present in comparable amounts. These forms can be experimentally observed by two-dimensional 1H15N NMR spectra of the solutions (natural 15N content is sufficient to obtain the spectra) if protons at the C5 and C6 atoms are not rapidly exchangeable with the solvent. Such spectra have been obtained in 2012 by Ilya El'tsov in the Novosibirsk State University (Novosibirsk, Russia) and can be reproduced by everybody who has NMR spectrometer and deoxycitidine solution. Taken this into account, one can expect presence of both tautomeric forms of cytosine derivatives in DNA as well.

At pH 7 spectra of deoxycitidine solutions contain signals of 5 various nitrogen atoms, while at pH=5 only 3 different nitrogen atoms signals are detectable. The latter effect takes place because at pH=5 in the solution protonated form of deoxycidtidine dominates, and in this form tautomeric transitions are impossible. Comparison of the tautomeric forms of unprotonated deoxycitidine with the protonated form (Fig. 1) shows that the amino group of the latter is analogous to that of canonical form of the unprotonated cytosine, and the imino group is similar to that of the imino-tautomer. At neutral pH the variety of nitrogen atoms increases since both amino- and imino groups can loss protons. Protonation should significantly affect chemical shift of only one nitrogen, therefore, the presence of the canonical amino-tautomer of cytosine and its protonated form should lead to only 4 different nitrogens in the NMR spectra. Observation of 5 different nitrogen signals unambiguously shows the presence of both cytosine tautomers in comparable amounts (which implies their comparable stabilities).

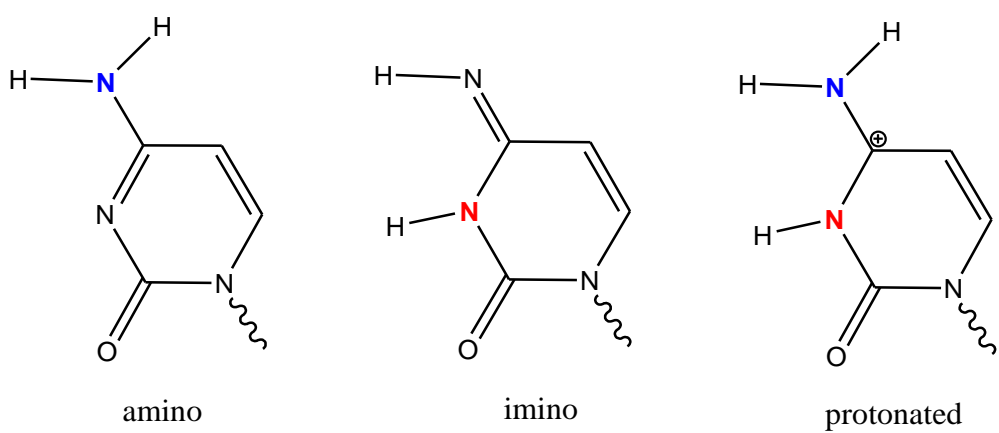

amino                imino                protonated

Figure 1. Protonated form of cytosine combines properties of nitrogen atoms of both amino- and imino-tautomers of the unprotonated form.

It is worth to note that with some cytosine derivatives, preferential formation of the "rare" imino tautomer is expected. For example, this can be the case with 5-methylcytosine because of the electron effects of the methyl group (Fig. 2A). Methyl group is electro donor at σ-bonds, which is expected to facilitate loss of proton from nitrogen of the amino group due to the following reasonings. In the amino tautomer of cytosine (Fig. 1), C4 neighbors weak electron acceptor N3 and such relatively strong donor as the amino group. In the imino tautomer, the weak donor (N3) and the strong acceptor (the imino group) make relative deficiency of electron density in the heterocycle. Methyl group partly compensate this deficiency and thereby stabilizes the imino tautomer of 5-methylcytosine.

With 5-hydroxymethylcytosine, the imino form can be additionally stabilized by intramolecular hydrogen bond (Fig. 2 B); imino form of 5-formylcytosine can be stabilized similar way, but in the case when the imino group is proton donor (Figure 2C).

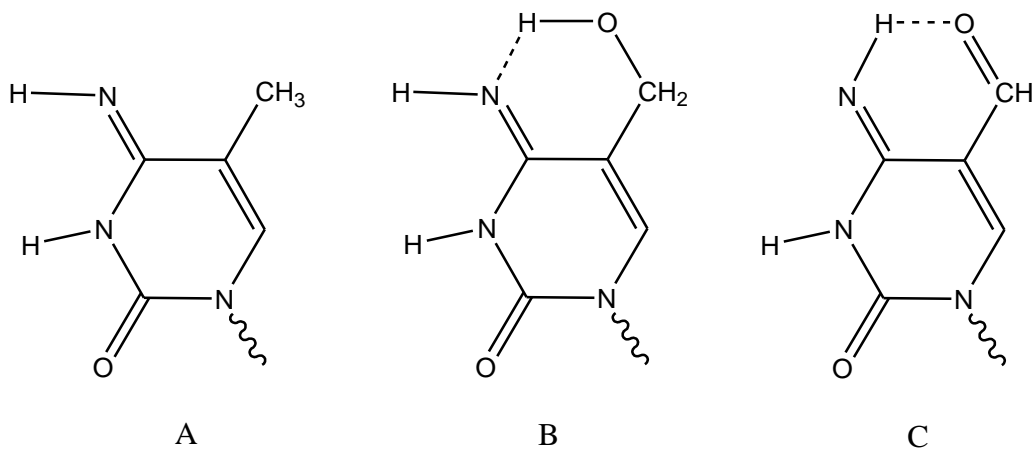

A  B  C

Figure 2. Imino-tautomer in various cytosine in various cytosine derivatives is stabilized: in 5-mehtylcytosine due to the electron effects of the methyl group (A), in 5-hydroxymethylcytosine because of the formation of intramolecular hydrogen bond (B) and in 5-formylcytosine due to the formation of hydrogen bond via the imino-proton (C).

Three above mentioned cytosine derivatives represent sequential steps of methylation-demethylation process underlying a well-known process of regulation of genes expression [1, 2]. In particular, it has been proved that methylation and hydroxymethylation of particular DNA cytosines greatly affects genes expression [2]. Notably, this is not the case with formylation of cytosine [2]. The goal of the further part of this article is to show how tautomerism can significantly affect epigenetic effects of cytosine derivatives.

Figure 3 shows a suggested base pair formed by the imino tautomer of cytosine and guanine in the geometry of GC-Hoogsteen base pair which has not been reported in literature so far but seems probable from structural point of view. Fig. 3a presents a structure built by analogy with well-known $GC^+$-Hoogsteen base pair (Fig. 3b) which is more expected when pH is <5. It is naturally to suggest that cytosine derivatives that have increased ability to form the imino tautomer can easier form the pair like that shown in Fig. 3. Fig. 4 shows that formation of the mentioned pair type is probable with 5-methylcytosine and 5-hydroxymethylcytopsine but impossible with 5-formylcytosine, which is in agreement with extent of participation of these cytosine derivatives in epigenetic processes [1].

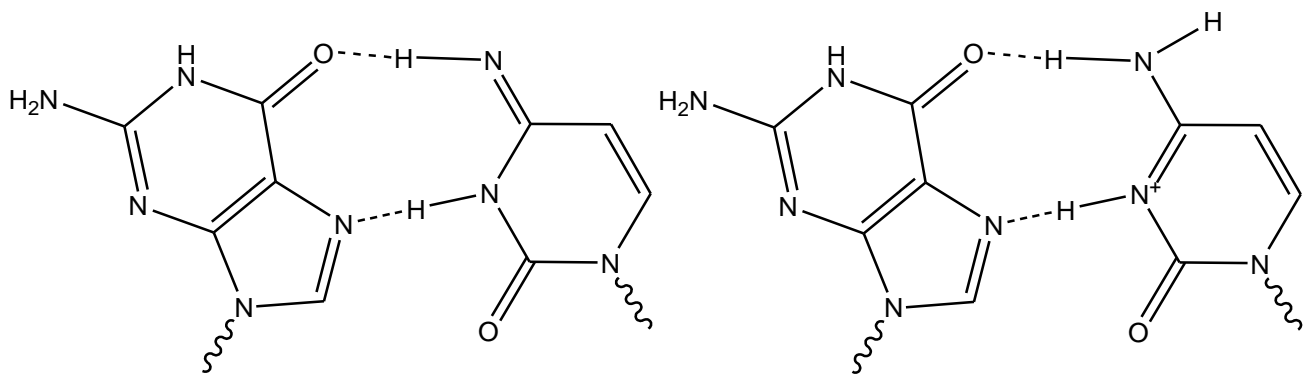

Figure 3. (a) GC-Hoogsteen pair formed by imino tautomer of cytosine as likely variant of the base pair formation at pH near 7. (b) Classical GC$^+$-Hoogsteen base pair

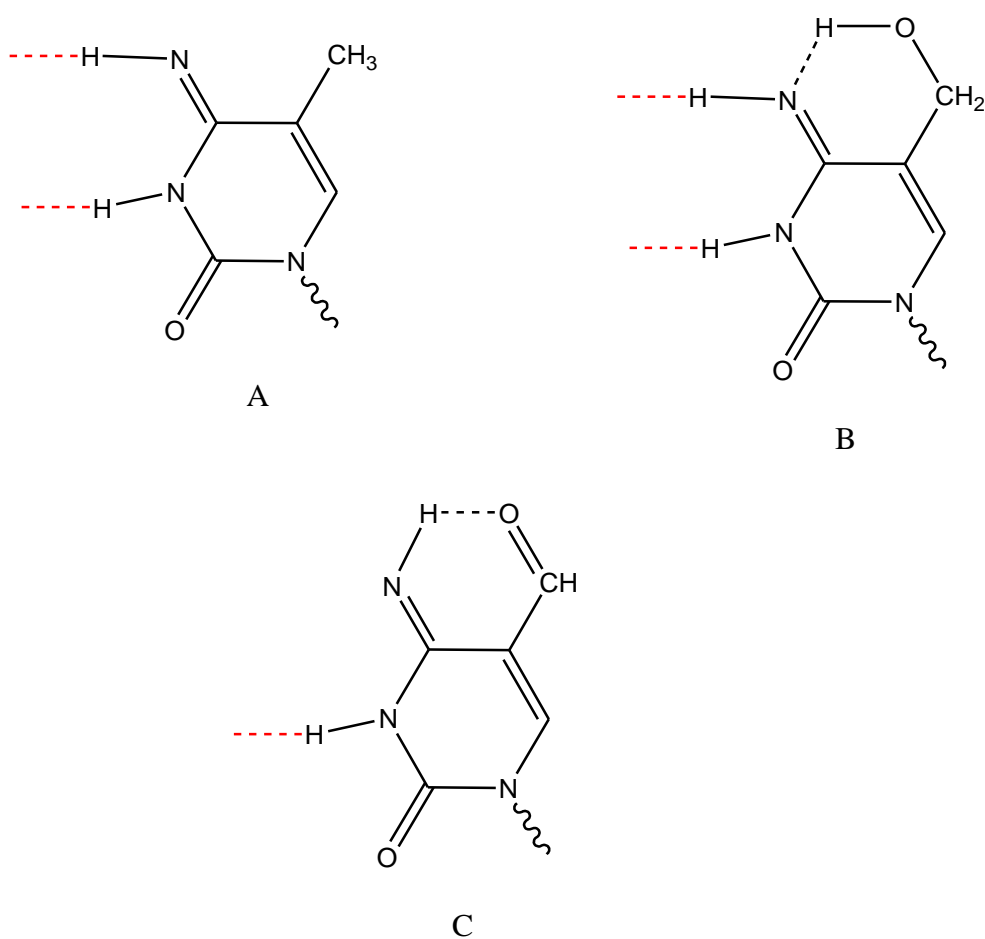

Figure 4. GC-Hoogsteen pair potentially could be formed with participation of 5-methyl- and 5-hydroxymethylcytosine but 5-formylcytosine is unable to form such type pair because of formation of the intramolecular hydrogen bond.

GC-Hoogsteen base pair was suggested to be the reason for epigenetic processes in the works of Al-Hashimi et al. [3, 4], where spontaneous formation of these pairs was demonstrated to occur in the DNA double helix. In these works only protonated GC$^+$-Hoogsteen base pair was considered since the work has been carried out at pH<5,5. Under these conditions, protonated form of cytosine dominates, and the presence imino-tautomer is unlikely. In general, the experiments under acidic conditions do not completely explain epigenetic properties of cytosine derivatives. Protonated forms of cytosine and all his derivatives mentioned above should not have significant dissimilarities upon formation of pairs similar to the GC$^+$-Hoogsteen base pair. It is worth to mention here that protonated form of 5-formylcytosine should be able to form GC+-Hoogsteen base pair. However, this cytosine derivative has no epigenetic effects, which indicates that these effects are not caused by protonated forms of cytosine derivatives.

One of methods used by Al-Hashimi et al. Could be applied for experimental testing of the suggestion on existence of GC-Hoogsteen pairs formed by imino tautomer of cytosine.To stabilize GC$^+$-Hoogsteen pairs in [3] 1-methylguanine was used, which prevented formation of the GC Watson-Crick base pair. At pH 7 this approach is expect to make possible detection of the suggested base pair (Fig. 6). To obtain a proof for the tautomeric structure of cytosine participating in the base pair formation, it is necessary to apply 1H15N NMR. In these experiments chemical shifts of the nitrogen atoms should exactly show which forms of cytosine are involved, protonated or tautomeric. Information concerning relative stability of the tautomers originated from various cytosine derivatives can be obtained from the comparison of Tm of the corresponding duplexes.

There is an experimental indication for the increased probability of tautomeric transition with 5-formylcytosine. The latter can form a stable complementary pair with adenine wobble position of codon [5]. Formation of the imino tautomer of 5-formylcytosine in this pair was shown by X-ray crystallography [5], and the geometry of this pair is similar to that of the AU pair (Fig. 5). One can expect that studying pair 5-formylC-A by NMR approaches will make it possible confirm the tautomer existence and besides provide data on intramolecular hydrogen bond that deprives imino tautomer of 5-formylcytosine from ability to form GC-Hoogsteen base pair.

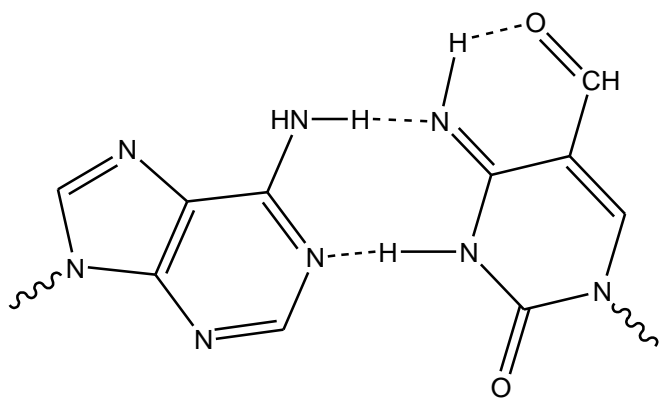

Figure 5. Pair adenine-5-formylcytosine formed by imino tautomer of the cytosine derivative. The tautomer can be stabilized by the intramolecular hydrogen bond that should be well recognizable on the NMR spectra.

**The main conclusion:** methylation and hydroxymethylation of cytosine can shift tautomer equilibrium to the imino tautomer, and the latter can form GC-Hoogsteen base pair at pH near 7, which can be the reason for epigenetic effects of 5-methylcytosine and 5-hydroxymethylcytosine.

**Acknowledgements:**
The author is grateful to D.M. Graifer for helpful discussion and his help in the text preparation.